\documentclass[pre,onecolumn,superscriptaddress,nofootinbib]{revtex4}

\usepackage{graphicx}
\usepackage{dcolumn}
\usepackage{array}
\usepackage{tabularx}
\usepackage{color}

\usepackage{floatrow}

\usepackage{eucal}
\usepackage[dvips]{epsfig}
\usepackage{amssymb}
\usepackage{amsfonts}

\usepackage{amsmath,amsthm}
\newcommand{\be}{\begin{eqnarray}}
\newcommand{\ee}{\end{eqnarray}}
\newcommand{\bse}{\begin{subequations}}
\newcommand{\ese}{\end{subequations}}

\newcommand{\bnum}{\begin{enumerate}}
\newcommand{\enum}{\end{enumerate}}

\newcommand{\bit}{\begin{itemize}}
\newcommand{\eit}{\end{itemize}}

\newcommand{\bc}{\begin{cases}}
\newcommand{\ec}{\end{cases}}

\newcommand{\bpm}{\begin{pmatrix}}
\newcommand{\epm}{\end{pmatrix}}

\newcommand{\bvm}{\begin{vmatrix}}
\newcommand{\evm}{\end{vmatrix}}

\newcommand{\bs}{\boldsymbol}

\newcommand{\mcal}{\mathcal}

\newcommand{\mrm}{\mathrm}

\newcommand{\gd}{\delta}
\newcommand{\eps}{\epsilon}%\ge schon vergeben

\newcommand{\gs}{\sigma}

\newcommand{\Gc}{\Gamma}

\newcommand{\Gl}{\Lambda}

\newcommand{\p}{\partial}
\newcommand{\f}{\frac}

\newcommand{\lan}{\langle}
\newcommand{\ran}{\rangle}

\begin{document}

%\title{Spontaneous flow and geometry-dependent viscosity reduction in sheared active fluids}
\title{Geometry-dependent viscosity reduction in sheared active fluids}

\author{Jonasz S\l{}omka}
\affiliation{Department of Mathematics, Massachusetts Institute of Technology, 77 Massachusetts Avenue, Cambridge, MA 02139-4307, USA}

\author{J{\"o}rn Dunkel}
\affiliation{Department of Mathematics, Massachusetts Institute of Technology, 77 Massachusetts Avenue, Cambridge, MA 02139-4307, USA}

\date{\today}
\begin{abstract}
We investigate flow pattern formation and viscosity reduction mechanisms in active fluids by studying a generalized Navier-Stokes model that captures the experimentally observed bulk vortex dynamics in microbial suspensions. We present exact analytical solutions including stress-free vortex lattices  and introduce a computational framework that allows the efficient treatment of previously intractable higher-order shear boundary conditions. Large-scale parameter scans identify the conditions for spontaneous flow symmetry breaking, geometry-dependent viscosity reduction and  negative-viscosity states amenable to energy harvesting in confined suspensions.  The theory uses only generic assumptions about the symmetries and long-wavelength structure of active stress tensors, suggesting that inviscid phases may be achievable in a broad class of non-equilibrium fluids by tuning confinement geometry and pattern scale selection.
\end{abstract}

%\end{abstract}

\maketitle

%%%%%%%%%%%%%%%%%%%%%%%%%%%%%%%%%%%%%%%

Self-driven vortical flows in microbial~\cite{1997Kessler} and synthesized active liquids~\cite{2014Keber_Science,2012Sanchez_Nature,2015DeCamp} often exhibit a dominant length scale~\cite{2004DoEtAl,2007SoEtAl,2012Wensink,2012Sokolov}, distinctly different from the scale-free spectra of conventional turbulence~\cite{2004Frisch}. Experimentally observed vortices in dense bacterial suspensions typically have diameters $\Gl\sim 50-100\,\mu$m~\cite{2004DoEtAl,2012Sokolov,2013Dunkel_PRL} and decay within a few seconds in a bulk fluid~\cite{2013Dunkel_PRL}.  However, when the suspension is enclosed by a small container of dimensions comparable to $\Gl$, individual vortices become stabilized for several minutes~\cite{2013Wioland_PRL,2014Lushi_PNAS} and can be coupled together to form magnetically ordered vortex lattices~\cite{2016Wioland_NPhys}. Another form of confinement-induced symmetry breaking was observed recently in a microfluidic realization of bacterial \lq racetracks\rq{}~\cite{2016Wioland_RaceTracks}. For sufficiently narrow tracks of diameter $\ll \Gl$, bacteria spontaneously aligned their swimming directions to form persistent unidirectional currents.  These examples illustrate the importance of confinement geometry for flow-pattern formation in non-equilibrium liquids. Conversely,  biologically or chemically powered fluids may profoundly affect the dynamics of moving boundaries as active components can significantly alter the effective viscosity of the surrounding solvent fluid~\cite{2009SkolovAranson,2010RaJiPe,2013Lindner}.  In particular, recent shear experiments suggest that  \textit{Escherichia coli}  bacteria can create effectively inviscid flow if their concentration and activity are sufficiently large to support coherent collective swimming~\cite{2015LoGaDoAuCle}. From a theory perspective, it is desirable to formulate a minimal hydrodynamic model that is analytically tractable and can account for all the aforementioned experimental observations without overfitting.
\par
\begin{figure*}[t!]
\centering
\includegraphics[width=1.0\columnwidth]{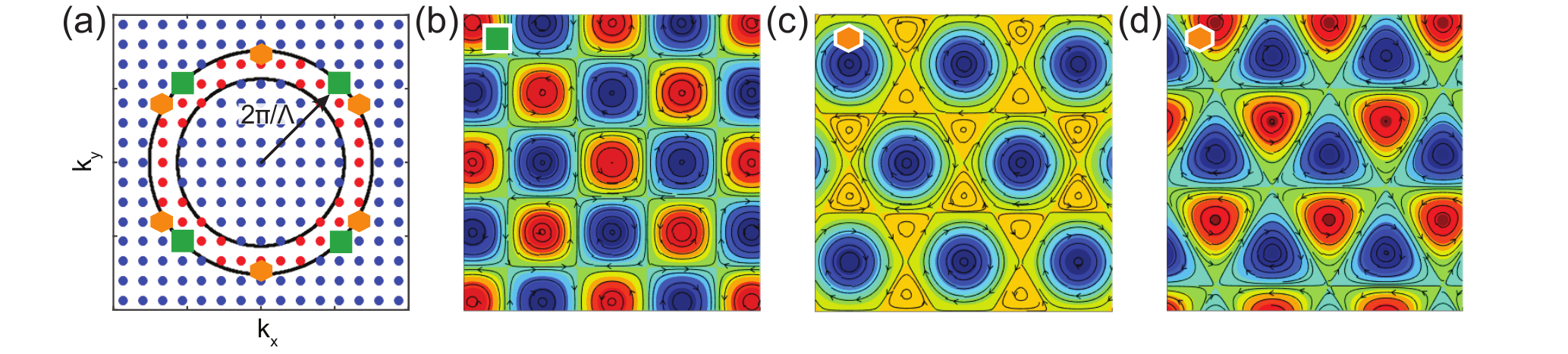}
 \caption{
Exact periodic solutions of Eqs.~(\ref{e:eomvsf}) include inviscid vortex lattices.
(a)~Linear stability analysis of Eqs.~\eqref{e:eomvsf}, for a square domain with periodic boundary conditions
identifies three types of Fourier modes: dissipative (blue), active (red), and neutral (black circles). Neutral modes can be combined through  the superposition~\eqref{e:exact_sol} to form exact stationary stress-free solutions of the nonlinear Eqs.~\eqref{e:eomvsf}. 
(b)~Square lattice solution, $\psi=\cos(\f{kx}{\sqrt{2}})\cos(\f{ky}{\sqrt{2}})$, corresponding to the green squares in panel~(a).
(c)~Hexagonal Kagome lattice solution, $\psi=2\cos(\f{ky}{2})\cos(\f{\sqrt{3}kx}{2})-\cos(k y)$,
corresponding to the orange hexagons in~(a). This flow topology is very similar to Abrikosov lattices found in quantum superfluids, cf. Figures in~\cite{1957Abrikosov,1964Kleiner}.
(d)~Triangular lattice solution, $\psi=-\cos(ky)-\cos(k\f{\sqrt{3}x-y}{2})-\sin(k\f{\sqrt{3}x+y}{2})$,
also corresponding to the orange hexagons in~(d), but with different mode amplitudes.
In all three cases, $k$ can be chosen as the radius of the inner or the outer inviscid circle (black) in panel~(a). 
}
\label{fig01} 
\end{figure*}

Previous theoretical work~\cite{2008CaFieMarOrYoe,haines2009three,giomi2010sheared,PhysRevE.83.041910,ryan2011viscosity,2012Foffano_EPJE} identified potential viscosity reduction mechanisms~\cite{2009SkolovAranson,2015LoGaDoAuCle} in certain classes of active suspensions, but the complexity and specific nature of the underlying multi-field models have made analytical insight, time-resolved dynamical studies and comparison with experiment challenging.  To better understand the general conditions under which active fluids can develop spontaneous symmetry-breaking and quasi-inviscid behavior, we pursue here an alternative approach by focusing on the generic  phenomenological properties of non-Newtonian fluids that exhibit biologically, chemically or physically driven pattern formation. Specifically, we derive exact 2D bulk solutions for a higher-order generalization of the classical Navier-Stokes (NS) equations (Fig.~\ref{fig01}) that accounts for the self-sustained vortex dynamics of bacterial suspensions seen in experiments~\cite{1997Kessler,2004DoEtAl,2007SoEtAl,2012Sokolov}.  In contrast to earlier studies of the bacterial velocity field~\cite{2012Wensink,2013Dunkel_PRL}, the discussion below focuses exclusively on the solvent flow dynamics relevant to shear experiments.  We complement our theoretical considerations with large-scale simulations, introducing  a  numerical framework that allows the efficient treatment of previously intractable higher-order shear boundary conditions.  Our results show that a two-parameter extension of the classical Navier-Stokes theory suffices to describe the recently reported spontaneous symmetry-breaking phenomena~\cite{2016Wioland_RaceTracks} and inviscid phases of bacterial suspensions~\cite{2015LoGaDoAuCle}. Furthermore, the theory yields testable predictions for viscosity resonances mediated by topological  defects in the stress field, and provides guidance for the optimal design of Taylor-Couette motors~\cite{furthauer2012taylor} powered by active meso-scale turbulence~\cite{2007SoEtAl,2012Wensink,2013Dunkel_PRL,Bratanov08122015,PhysRevLett.112.158101}. Generally, our analysis suggests that low-viscosity modes may be generically present in a wide range of chemically or biologically driven~\cite{2008SaintillanShelley} fluids, and that such modes can be selected and exploited by an optimal tuning of active vortex scales and boundary geometry. 

%%%%%%%%%%%%%%%%%%%%%%%
%%%%%%%%%%%%%%%%%%%%%%%

%%%%%%%%%%%%%%%%%%%%%
\section*{THEORY}
%%%%%%%%%%%%%%%%%%%%%

%%%%%%%%%%%%%%%%%%%%%
\textbf{Generalized Navier-Stokes model}.
We describe the incompressible solvent flow field $\bs v(t,\bs x)$ in the presence of microorganisms or other active components  
by the NS equations
\bse
\label{e:eom}
\be
\label{e:eoma}
\nabla\cdot\bs v&=&0, \\
\label{e:eomb}
\p_t \bs v+\bs v \cdot \nabla \bs v&=&-\nabla p+\nabla \cdot{\bs \sigma},
\ee
\ese
where   $p(t,\bs x)$ denotes the local pressure.  The effective stress tensor $\gs(t,\bs x)$ comprises passive contributions from the intrinsic fluid viscosity and  active contributions representing the forces exerted by the microswimmers on the fluid~\cite{2002Ra,2013Marchetti_Review,2013Ravnik_PRL}. At sufficiently high concentrations,  bacteria swimming collectively through an ambient fluid create vortices of typical diameter $\Lambda\sim 50-100\mu$m~\cite{2012Sokolov,2013Dunkel_PRL}. Although the microscopic origins of self-organized scale-selection in active fluids are not yet fully understood, one can model the experimental observations phenomenologically through the stress tensor~\cite{2015SlomkaDunkel}
\be
\label{e:stress}
\bs\gs=(\Gamma_0 -\Gamma_2 \nabla^2+\Gamma_4 \nabla^4)[\nabla \bs v+ (\nabla \bs v)^\top],
\ee
where the higher-order derivatives $\nabla^{2n}\equiv (\nabla^2)^n$, \mbox{$n\ge2$} account for non-Newtonian effects~\cite{2014BipolarBook}. The one-dimensional version
of Eqs.~\eqref{e:eom} and~\eqref{e:stress} is also known as the generalized Nikolaevskiy model~\cite{1993BeNi_PhysD} and has been studied in the context of soft-mode turbulence and nonlinear seismic waves~\cite{1996Tribelsky_PRL,PhysRevE.77.035202}.

\par
Intuitively,  Eq.~\eqref{e:stress} is obtained by truncating a long-wavelength expansion of the (unknown) full stress-tensor~\cite{2014BipolarBook}. For $\Gc_2=\Gc_4=0$, Eqs.~\eqref{e:eom} and \eqref{e:stress} reduce to the standard NS equations of a passive fluid with kinematic viscosity $\Gamma_0>0$.  For $\Gc_0>0,\Gc_4>0$ and $\Gc_2<0$, the ansatz~\eqref{e:stress} is the simplest choice of an active stress tensor that is isotropic, selects vortices of a characteristic scale, and yields a stable theory at small and large wave numbers~\cite{2015SlomkaDunkel}. The transition from an active to a passive fluid, which can be realized experimentally through ATP or nutrient depletion, corresponds to a sign change from $\Gamma_2<0$ to $\Gamma_2\ge 0$, whereas the non-negativity of  $\Gamma_0$ and $\Gamma_4$ follows from general stability considerations.
\par
For \lq scale-free\rq{} passive  Newtonian fluids, $\Gc_0$ encodes collective molecular interactions and thermal effects, while higher-order effects can typically be neglected. For  pattern-forming active fluids, the effective parameters $(\Gamma_0,\Gamma_2,\Gamma_4)$ contain contributions from microscopic interactions, thermal and athermal fluctuations, and other non-equilibrium processes. In this case, $\Gamma_0$ describes the damping of long-wavelength  perturbations  on scales much larger than the typical correlation length of the coherent flow structures, whereas $\Gamma_{2}$ and $\Gamma_4$ account for the growth and damping of modes at intermediate and small scales. For suitably chosen values of $(\Gamma_0,\Gamma_2,\Gamma_4)$,  Eqs.~\eqref{e:eom} and \eqref{e:stress} reproduce the experimentally observed bulk vortex dynamics of bacterial suspensions~\cite{2007SoEtAl,2012Sokolov,2013Dunkel_PRL} (Fig.~\ref{fig02}). These non-equilibrium flow structures can be characterized in terms of the typical vortex size $\Lambda=\pi\sqrt{2\Gamma_4/(-\Gamma_2)}$, growth timescale $\tau=\tau(\Gamma_0,\Gamma_2,\Gamma_4) $  (App.~\ref{a:Characteristic scales}) and circulation speed $U=2\pi\Lambda/\tau$. For example, the parameter choice $\Gamma_0=10^3\,\mu$m$^2$/s, $\Gamma_2/\Gamma_0=-1.24\times10^2\,\mu$m$^2$, $\Gamma_4/\Gamma_0=3.53\times 10^4\,\mu$m$^4$ yields values $\Lambda=75\,\mu$m, $\tau=6.6\,$s and $U=72\,\mu$m/s that match well recent measurements for \textit{Bacillus subtilis} suspensions~\cite{2007SoEtAl,2012Sokolov,2013Dunkel_PRL}. 
\par
More generally, however, truncated polynomial stress-tensors of the form~\eqref{e:stress} can be expected to provide useful long-wavelength approximations for a broad class  of pattern-forming liquids,  including magnetically~\cite{2008Ouellette}, electrically~\cite{C5SM02316E}, thermally~\cite{PhysRevLett.105.268302,2014Bregulla_ACSNano,2015Fedosov_SoftMatter} or chemically~\cite{2013Loewen_PRL,2013Buttinoni_PRL} driven flows. Below, we show that  pattern-forming active suspensions described by Eqs.~\eqref{e:eom} and \eqref{e:stress} can exhibit frictionless and negative-viscosity dynamics.

%%%%%%%%%%%%%%%%%%%%%
%\newpage
\textbf{2D vorticity-stream function formulation}.
The generalized NS equations~\eqref{e:eom} and~\eqref{e:stress} are valid in arbitrary dimensions. Here, we focus on the 2D case relevant to free-standing\footnote{To describe thin-film experiments performed on a substrate, one could add a linear damping term $-\gamma_0\bs v$ in the NS equations to account phenomenologically for the substrate friction. However, such a modification would merely lead to a trivial shift of the dispersion relation. Therefore, if the damping is not supercritical and active vortical flows are not completely suppressed, then one can expect that the main results of this study remain valid qualitatively for films on substrates.}  thin-film experiments~\cite{2007SoEtAl}. In a planar 2D geometry $\mcal{D}$ with boundary $\p\mcal{D}$, we may rewrite Eqs.~\eqref{e:eom} and~\eqref{e:stress} in vorticity-stream function form  (App.~\ref{a:hodge})
\bse
\label{e:eomvsf}
\be
\label{e:eomvsfa}
\p_t \omega+\nabla\omega\wedge\nabla\psi
&=&-\bs H\cdot\nabla\omega +
%\\&&\notag
\Gamma_0\nabla^2\omega-\Gamma_2\nabla^4\omega+\Gamma_4\nabla^6\omega, 
\qquad\\
\label{e:eomvsfb}
\nabla^2\psi&=&-\omega,
\ee
\ese
where the vorticity $\omega=\nabla\wedge \bs v=\eps_{ij}\partial_iv_j$  is defined in terms of the 2D Levi-Civita tensor $\eps_{ij}$, $\psi$ is the stream function and $\bs H$ is a harmonic field related to the fluid's center-of-mass (CM) motion. The components of the flow field  $\bs v=(v_1,v_2)$ are recovered from the Hodge decomposition \cite{schwarz1995hodge}  as $v_i=\eps_{ij}\p_j \psi+H_i$   (App.~\ref{a:hodge}). 
\par

\textbf{Analytical solutions \& zero-viscosity modes}.
We construct a family of exact nontrivial stationary solutions of the nonlinear partial differential equations (PDEs)~\eqref{e:eomvsf} in free space. To this end, we focus on the center-of-mass  frame with $\bs H=0$ and consider the stream-function ansatz
\be
\label{e:exact_sol}
\psi(r,\theta)=\int_0^{2\pi}d\phi\; \hat \psi(\phi)\,e^{-ikr \cos(\phi-\theta)},
\ee
where $k=\sqrt{k_x^2 +k_y^2}$ is a fixed wavenumber radius, and  $(r,\theta)$ are polar position coordinates. The superposition~\eqref{e:exact_sol} yields the vorticity 
\be
\omega=-\nabla^2 \psi=k^2 \psi,
\ee 
and hence eliminates the nonlinear advection term in Eq.~\eqref{e:eomvsfa}, because 
\be
\nabla \omega\wedge \nabla \psi=k^2\nabla \psi\wedge \nabla \psi=0.
\ee 
Thus, to obtain a stationary solution of Eqs.~\eqref{e:eomvsf}, we need to fix $k$ such that the rhs. of Eq.~\eqref{e:eomvsfa} vanishes. This criterion can be fulfilled if $k$ satisfies the polynomial equation 
\be
k^2(\Gamma_0+\Gamma_2 k^2+\Gamma_4 k^4)=0,
\ee 
which has real roots if $\Gamma_2<0$ and $\Gamma_2^2>4 \Gamma_0\Gamma_4$. 
\par
One can further show that the stress tensor defined in Eq.~\eqref{e:stress} vanishes identically, $\bs \sigma\equiv 0$, for stationary solutions of this type. Thus, these solutions are stress-free  modes, describing effectively frictionless flow states (Fig.~\ref{fig01}a). An interesting subclass of exact stationary solutions included in Eq.~\eqref{e:exact_sol} are vortex lattices. By superimposing a small number of $k$-modes that lie on one of the two stress-free rings, with $\hat\psi$ being a sum of suitably weighted Dirac delta-functions, one can  construct rectangular, hexagonal and triangular lattices (Fig.~\ref{fig01}b-d), whereas oblique lattices are forbidden by rotational symmetry. The stress-free solutions lie at the interface of the stable and unstable modes (Fig.~\ref{fig01}a). We next demonstrate through simulations that effectively inviscid behavior remains observable in shear experiments for optimized geometries.

%\pagebreak

%%%%%%%%%%%%%%%%%%%%%%%
\section*{SIMULATIONS}
%%%%%%%%%%%%%%%%%%%%%%%

\begin{figure*}[t!]
\centering
\includegraphics[width=\textwidth]{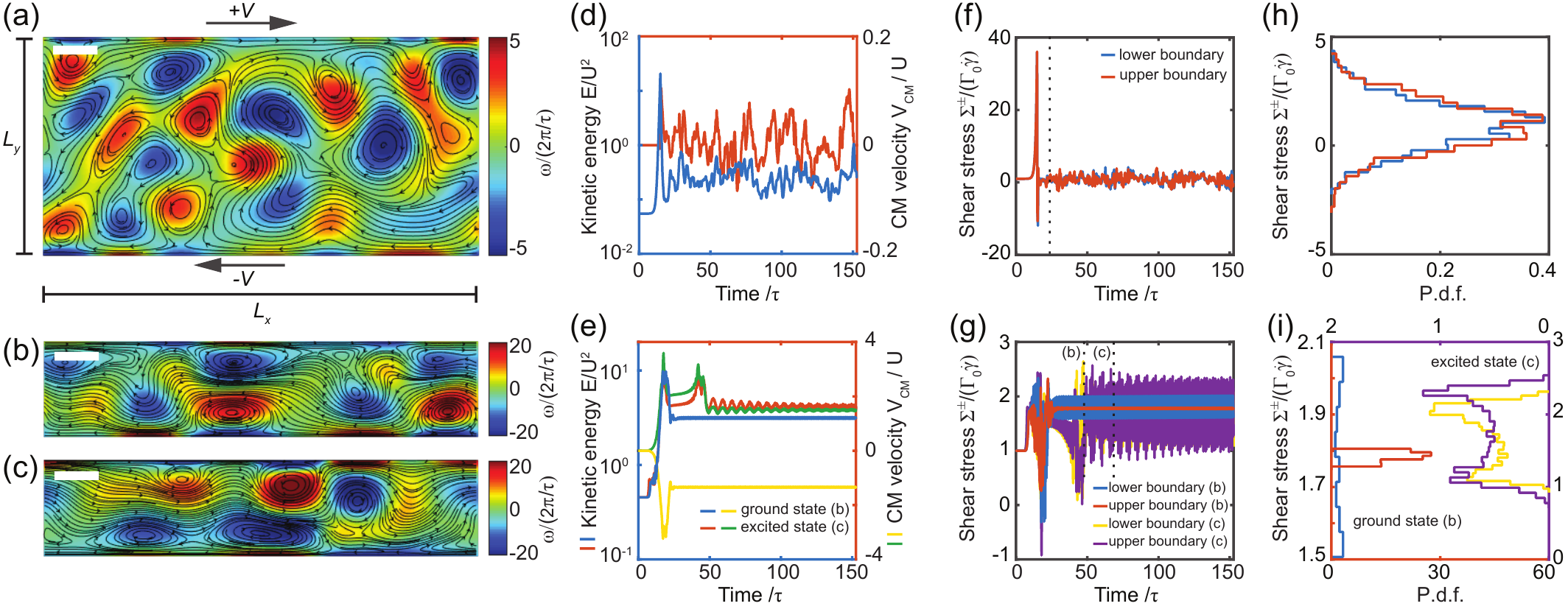}
 \caption{
Active shear flows exhibit qualitatively different dynamics, velocity statistics and symmetry-breaking behavior depending on confinement geometry $(L_x,L_y)$ and applied shear speed~$V$ (see also Fig.~\ref{fig03}).
(a), 
For wide channels of width $L_y=5\Gl$ and weak shear $V=0.57U$, typical flow configurations realize advectively mixed vortex lattices~(Movie~1).  The characteristic circulation speed $U$ and growth-time scale $\tau$ of the bulk vortices are defined in App.~\ref{a:Characteristic scales}. Scale bar shows bulk vortex size $\Lambda$ and colormap encodes vorticity. 
 (b-c), 
 For narrow channels with $L_y=2.2\Lambda$ and strong shear $V=1.65U$, 
the active fluid exhibits spontaneous symmetry-breaking resulting in unidirectional transport of the fluid's center of mass (Movies~2,3 and Fig.~\ref{figS2}). Depending on initial conditions, qualitatively different low-energy (b; Movie~2) and high-energy (c; Movie~3) flow states can arise for identical system parameters. 
(d-e), 
The kinetic energy time series $E(t)$ for the simulations in panels (a-c) illustrate relaxation to statistically stationary states. The center-of-mass velocity $V_{\textnormal{CM}}(t)$ indicates persistent macroscopic average flows through the channel (Movies~2, 3 and Fig.~\ref{figS2}).
(f-g),
The spatially averaged mean shear stresses  $\Sigma^\pm(t)$, rescaled by kinematic viscosity $\Gamma_0$ and shear rate $\dot{\gamma}$,  reveal top-bottom symmetry-breaking in narrow channels (b,c,g) and exhibit large temporal variability, resulting in a substantial variance of the effective shear viscosity (see also Fig.~\ref{figS4}). Vertical dotted lines indicate the start time $T$ of the temporal averaging periods for the results depicted in Fig.~\ref{fig03}a-c.
(h-i),
Shear stress histograms constructed from the time series in (f-g) for $t>T$ reflect the top-bottom flow asymmetry in narrow channels.
\label{fig02}} 
\end{figure*}

\textbf{Numerical shear experiments.} 
To study the rheology of Eqs.~\eqref{e:eomvsf}, we simulate a typical shear experiment~\cite{2015LoGaDoAuCle} in which two parallel boundaries are moved in opposite directions, both at a constant speed $V$ (Fig.~\ref{fig02}a-d). Specifically, we consider a rectangular domain $(x,y)\in \mcal{D}=[-L_x/2,L_x/2]\times [-L_y/2,L_y/2]$ with periodic boundary conditions in the $x$-direction and non-periodic shear boundary conditions in the $y$-direction (Fig.~\ref{fig02}a).  In this case, the harmonic field $\bs H(t,\bs x)=V_\mrm{CM}(t)\hat{\bs x}$  coincides with the center-of-mass velocity and, hence, is  governed by Newton's force-balance law, where the force acting on the fluid is obtained by integrating the stress tensor $\bs \sigma$ over the boundary (App.~\ref{numerical_methods}). 
\par
As common in the shear analysis of passive fluids~\cite{2007Lauga_NoSlip}, we assume no-slip boundary conditions for the flow field~$\bs v(x,\pm L_y/2)=(\pm V,0)$, which translate into an overdetermined system~\cite{Quartapelle1981}  for the stream function~$\psi$ (App.~\ref{numerical_methods}). In contrast to the classical second-order NS equations, the sixth-order PDE~\eqref{e:eomvsfa} requires additional higher-order boundary conditions to specify solutions. Active components in a fluid can form complex boundary-layer structures~\cite{2013Wioland_PRL,2014Lushi_PNAS,2016Wioland_NPhys}, which are poorly understood experimentally and theoretically. To identify physically acceptable boundary conditions,  we tested different types of higher-order conditions. These test simulations  showed that imposing  $\nabla^2\omega=0$ and $\nabla^4\omega=0$  at the boundaries reproduces the vortical bulk flow patterns observed in free-standing thin bacterial films~\cite{2007SoEtAl}, whereas stiffer boundary conditions generally do not produce the experimentally observed flow structures. We therefore fix $\nabla^2\omega=0$ and $\nabla^4\omega=0$ at the upper and lower boundaries throughout this study.
 \par
 Numerical solution of the coupled nonlinear sixth-order PDEs~\eqref{e:eomvsf} with non-periodic boundary conditions for experimentally relevant domain sizes~\cite{2004DoEtAl,2007SoEtAl,2012Wensink,2013Dunkel_PRL} is computationally challenging. We developed an algorithm that achieves the required numerical accuracy by combining a well-conditioned Chebyshev-Fourier spectral method  \cite{2013TownsendOlver,townsend2014automatic} with a third-order semi-implicit time-stepping scheme~\cite{ascher1995implicit} and integral conditions for the vorticity field~\cite{Quartapelle1981} (App.~\ref{numerical_methods}). This novel computationally efficient   code, which runs in real-time on conventional CPUs,  can be useful in simulations of a wide range of fluid-based pattern-formation processes, including Kolmogorov flows~\cite{2008Ouellette}.

%%%%%%%%%%%%%%%%%%%%%%%
%%%%%%%%%%%%%%%%%%%%%%%

\textbf{Parameters \& observables.}
We performed systematic large-scale parameter scans of realistic bulk coefficients $(\Gamma_0,\Gamma_2,\Gamma_4)$ and boundary conditions $(\dot\gamma,L_x,L_y)$, where \mbox{$\dot{\gamma}={V}/{L_y}$} is the shear rate (Fig.~\ref{fig02}a-c). 
Non-dimensionalization reduces the effective number of parameters to four, which we chose to be $(\Gamma_2,\dot\gamma,L_x,L_y)$. We explored $>$200 experimentally relevant parameter combinations in total. For a given parameter set, we repeated numerical shear experiments at least 10 times,  initializing simulations with a randomly perturbed linear shear profile (App.~\ref{numerical_methods}). For each simulation, we recorded the spatial averages of the kinetic energy (Fig.~\ref{fig02}d,e)
\be
E(t)=\f{1}{L_xL_y}\int_\mcal{D} dx dy\, (\bs v^2/2)
\ee 
 and the kinematic shear stresses 
 \be
 \Sigma^\pm(t)=\f{1}{L_x}\int_{\p\mcal{D}^{\pm}} dx\, \gs_{yx}
 \ee acting on the top and bottom boundaries (Fig.~\ref{fig02}f-i).  The statistics of these time series are analyzed for an interval $[T, T+\Delta]$, where $T$ is chosen larger than the numerically determined flow relaxation time. The averaging interval $\Delta$ is taken sufficiently long to ensure convergence of statistical observables (Fig.~\ref{fig02}f,g, Fig.~\ref{figS1}a-e).  For each time series $\mcal{O}(t)$, we compute  mean values 
 \be
\label{eq:effshearvisc}
\lan \mcal{O}\ran &=&
\lim_{T,\Delta\to \infty}\f{1}{\Delta}\int_{T}^{T+\Delta}dt\; \mcal{O}(t)
\ee
and histograms  (Fig.~\ref{fig02}h,i), by performing additional ensemble averaging over simulation runs with identical parameters  but different initial conditions (Fig.~\ref{figS1}a-d). Of particular interest for the subsequent analysis are measurements of the total shear stress on the two boundaries, \mbox{$\lan \Sigma \ran = \lan \Sigma^+\ran +\lan\Sigma^-\ran$}, and the associated mean  kinematic viscosity 
\be
\nu=\lan \Sigma \ran/\dot{\gamma}.
\ee

%\pagebreak
\section*{RESULTS}

\begin{figure*}[t]
  \caption{
Numerical simulations of Eq.~\eqref{e:eomvsf} identify the conditions for spontaneous left-right symmetry-breaking and low-viscosity states.
(a),~ The relative mean kinetic energy  of the fluid's center-of-mass signals spontaneous symmetry-breaking.  Averages over 10 simulation runs for each of the 208 simulated parameter pairs   (markers) were connected using spline interpolations. Black dots: simulations settle into a single class of statistically stationary kinetic energy ground-states. Red circles: In addition to the ground-state, long-lived excited states are observed for randomly sampled initial conditions (App.~\ref{numerical_methods}). Gray triangles indicate the occurrence of dynamical symmetry-breaking characterized by uni-directional fluid transport with mean speed $>0.25 U$ persisting over time-scales $>100\tau$. 
 $\Lambda$ and $\tau$ are the characteristic bulk vortex size and the characteristic timescale in an equivalent system with periodic boundary conditions. 
The black diagonal line marks the corresponding characteristic flow speed $U$, separating the low-shear from the high-shear regime.  
(b),~The dark blue domains in the effective viscosity phase diagram correspond to quasi-inviscid parameter regimes.
(c),~Vertical cut through panel~(b) at constant shear rate $\dot{\gamma}=2.7\tau^{-1}$ showing oscillatory behavior of the shear viscosity with boundary separation $L_y$. Viscosity fluctuations are maximal when an integer number $n$ of vortices fits between the boundaries, $L_y\approx n\Lambda$.
(d),~Horizontal cut through panel~(b) at constant separation $L_y=3\Lambda$, illustrating the suppression of viscosity fluctuations at high shear (see also Fig.~\ref{figS4}).
(e-f),~Representative flow fields, with local speed (top) and vorticity (center) shown as background, and corresponding stress fields (bottom) in the low-viscosity regime~(e; Movie~4) and the high-viscosity regime~(f; Movie~5). These simulations were performed at the same shear rate but different boundary separations, as indicated in panels (b,c). The spectral norm $||\bs \sigma||_2$, corresponding to the largest eigenvalue of the stress tensor $\bs \sigma$, and the associated director field reveal the presence of zero-stress defects in the bulk as well as half-loops in the stress-field lines along the edges for the low-viscosity states (e, bottom; Movie~4).
\label{fig03}
} 
  \centering
    \includegraphics[width=0.95\textwidth]{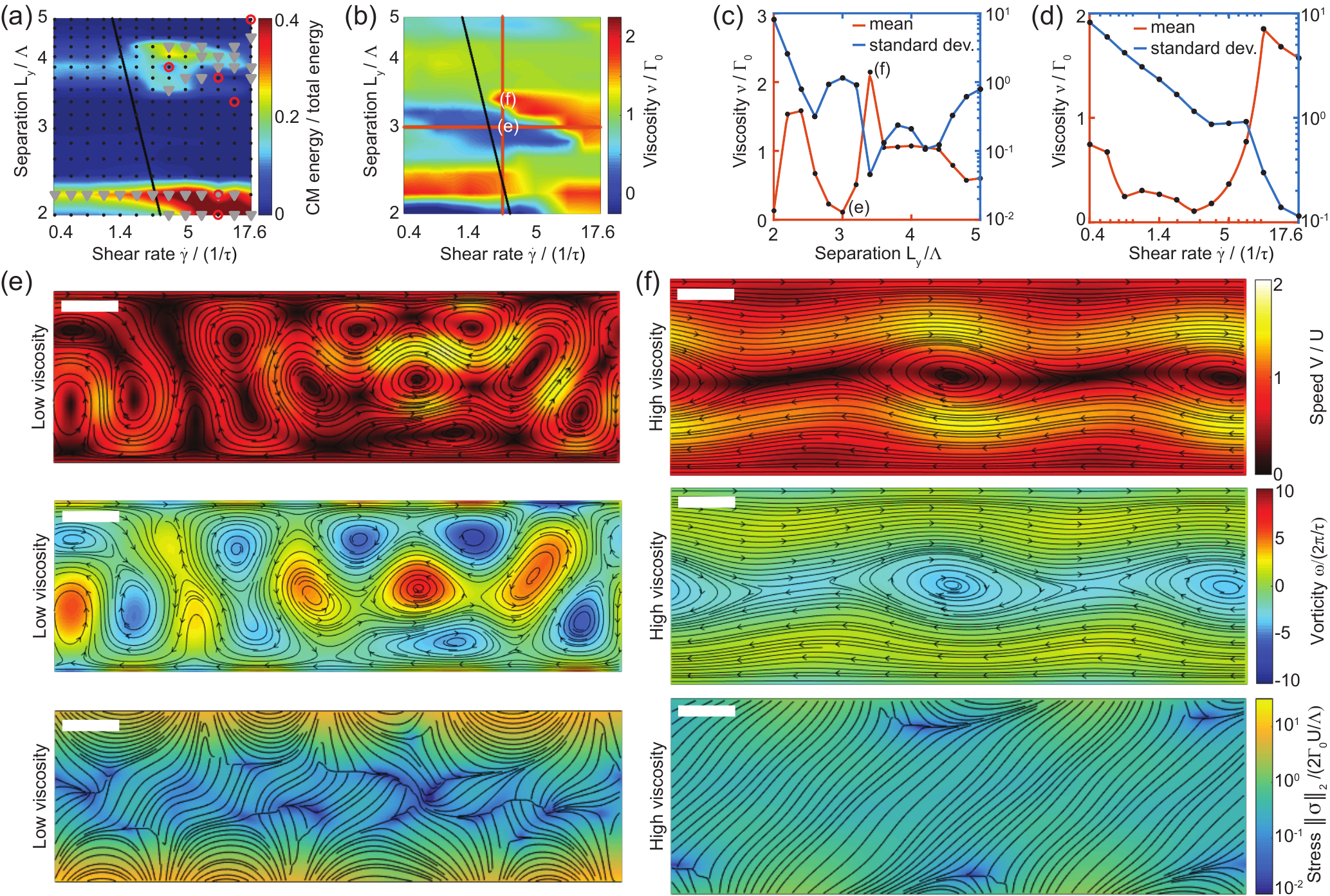}
\end{figure*}

\textbf{Dynamic symmetry breaking and directed transport.}
Recent experimental studies of bacterial suspensions~\cite{2016Wioland_RaceTracks} and ATP-driven active liquid crystals~\cite{DogicPrivate} in long narrow channels observed the spontaneous formation of persistent unidirectional macro-scale flows~\cite{2013Ravnik_PRL,2012Woodhouse}.  Our generalized NS model reproduces this dynamical symmetry-breaking effect (Fig.~\ref{fig02}, Fig.\ref{figS3}) and predicts optimal geometries that maximize directed transport (Fig.~\ref{fig03}a).  Fixing  $\Gamma_2<0$  to realize bacterial vortex structures as described above, we investigate how the boundary separation $L_y$ and the shear rate $\dot\gamma$ affect  mean velocity $V_\text{CM}$ of an active fluid modeled by Eqs.~(\ref{e:eom}). 
For wide channels with $L_y\gg \Gl$, the flow structures found in the simulations typically resemble a mixed vortex lattice (Fig.~\ref{fig02}a). In this case, the mean flow can fluctuate but is typically undirected  (Fig.~\ref{fig02}d; Movie~1). By contrast, for narrow channels, the center-of-mass velocity can spontaneously select a persistent mean-flow direction (Fig.~\ref{fig02}b,c,e; Movies~2,3). Our parameter scans show that this  
broken-symmetry phase extends over a wide range of shear rates if approximately two ($L_y\sim 2\Gl$) or four ($L_y\sim 4\Gl$) rows of vortices fit between the boundaries (Fig.~\ref{fig03}a). These results are in good qualitative agreement with recent microfluidic measurements in linearly confined bacterial suspensions; cf. Fig.~4 in Ref.~\cite{2016Wioland_RaceTracks}.

\textbf{Frustrated vortex packings.}
In addition to unidirectional center-of-mass motions, our simulations predict another 
secondary top-bottom symmetry-breaking phenomenon.
When the boundary separation is close to $2\Lambda$, the stress statistics for the two boundaries can be substantially different at high shear $V>U$ (Fig.~\ref{fig02}g,i).  Intuitively, this statistical  asymmetry can be explained by the fact that two counter-rotating vortices cannot simultaneously satisfy the externally imposed shear boundary conditions. Thus, one of the two vortices will be effectively pushed away from the boundary. The resulting asymmetric vortex alignment produces unequal shear forces on upper and lower boundaries even after long-time averaging  (Fig.~\ref{fig02}i), illustrating that the rheological analysis of active fluids requires more sensitive measures than in the case of passive fluids. 

\par
\textbf{Low-viscosity phases and edge-stresses.}
Recent experiments~\cite{2015LoGaDoAuCle} reported the observation of zero- and negative-viscosity states in concentrated~\textit{Escherichia coli} suspensions. Adopting typical bacterial parameters  $(\Gl,\tau,U)$ as described above, we investigate how the boundary separation $L_y$ and the shear rate $\dot\gamma$ affect the effective viscosity $\nu$ in the general NS model (Fig.~\ref{fig03}b-f).  Consistent with the experimental observations~\cite{2015LoGaDoAuCle}, the numerically obtained $(\dot\gamma, L_y)$-phase diagram confirms the existence of an effectively inviscid phase with $\nu/\Gamma_0\ll 1$ at low-to-intermediate values of the shear rate~$\dot\gamma$, when the boundary separation is around $3\Lambda$ (blue domain in Fig.~\ref{fig03}b). Varying the shear rate $\dot\gamma$ at constant separation $3\Lambda$, one observes a viscosity minimum when $\dot\gamma$ matches approximately the inverse vortex growth rate~$1/\tau$  (Fig.~\ref{fig03}d). In this quasi-inviscid regime, three counterrotating vortices fit between the boundaries, so that  the flow near the top and bottom aligns optimally with the boundary velocity (Fig.~\ref{fig03}e, top;  Movie~4). The nematic field lines of the associated stress field \eqref{e:stress} are defined by the eigenspace axis of the largest eigenvalue $||\bs\sigma||_2$. In the low-viscosity state, these director field lines connect primarily to the same boundary,  and they are separated by stress-free defects concentrated in the bulk region (Fig.~\ref{fig03}e, bottom;  Movie~4). Thus,  only a few stress-carrying strings connect the two boundaries, resulting in a significantly reduced shear viscosity. 

\par
\textbf{Viscosity resonances.}
In contrast to a passive Newtonian fluid, the effective viscosity  $\nu$ of the active fluid generally depends nonlinearly on both the shear rate $\dot\gamma$ and boundary separation~$L_y$ (Fig.~\ref{fig03}b-d). Qualitatively, we can distinguish between two characteristic regimes, corresponding  to shear speeds $V=\dot\gamma L_y$ larger or smaller than the characteristic bulk vortex speed $U$ (black lines in Fig.~\ref{fig03}a,b). At small shear speeds, $V<U$, the effective viscosity $\nu$ and its fluctuations depend primarily on the boundary separation~$L_y$, exhibiting oscillatory behavior as $L_y$ increases (Fig.~\ref{fig03}a,b). Viscosity minima occur at selected integer multiples of the characteristic bulk vortex size $\Lambda$ and are separated by maxima that can exceed $\Gamma_0$ by more than a factor 2 (Fig.~\ref{fig03}c). In such high-viscosity states, the stress field is nearly defect-free and similar to that of a laminar Newtonian fluid, with most of the stress field lines connecting the two boundaries (Fig.~\ref{fig03}f, bottom; Movie~5).  At supercritical shear speeds, $V>U$, the viscosity $\nu$ depends on both $L_y$ and $\dot\gamma$, and viscosity fluctuations decrease strongly with~$\dot{\gamma}$, signaling that the bulk dynamics becomes dominated by the no-slip boundary conditions at high shear (Fig.~\ref{fig03}b,d).

%%%%%%%%%%%%%%%%%%%%%%%
%%%%%%%%%%%%%%%%%%%%%%%

%%%%%%%%%%%%%%%%%%%%%%%
\section*{DISCUSSION}
%%%%%%%%%%%%%%%%%%%%%%%

\begin{figure}[t!]
  \centering
    \includegraphics[width=0.60\textwidth]{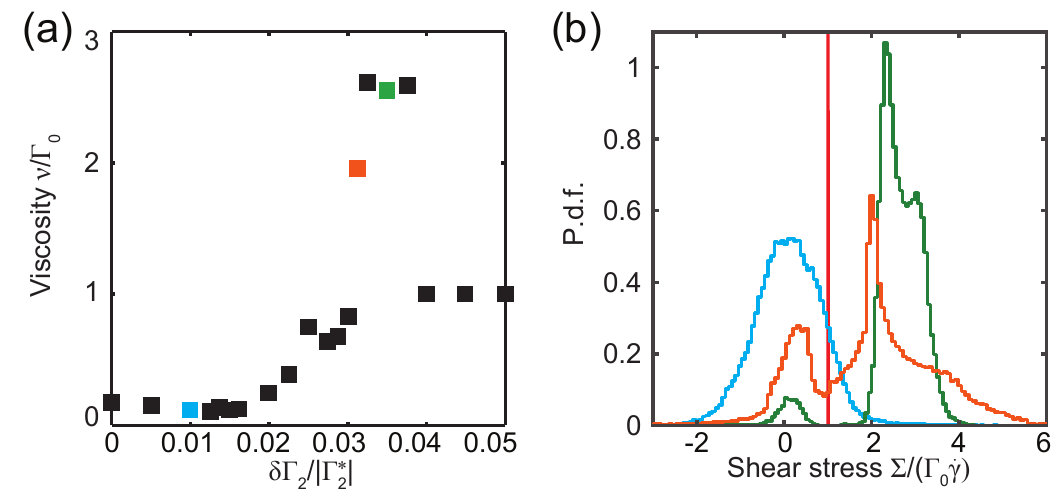}
  \caption{Transition from a low-viscosity to a normal fluid by changing the activity parameter $\Gamma_2=\Gamma_2^*+\gd \Gamma_2$,  starting from the quasi-inviscid state with $\Gamma_2^*<0$ in Fig.~\ref{fig03}e and keeping $L_y=3\Lambda$ and $\dot{\gamma}=2.7\tau^{-1}$ fixed. (a),~Increasing $\Gamma_2$ via  $\gd\Gamma_2$ corresponds to an effective reduction in activity.  As the activity is decreased, the effective shear viscosity  first increases before dropping to the value $\nu/\Gamma_0=1$, expected for a passive fluid with kinematic viscosity~$\Gamma_0$. 
(b),~Shear-stress histograms for the colored points in (a) show the transition from low-viscosity  flow (blue) to normal laminar flow (red) through a highly viscous state (green). In the vicinity of the critical point (orange) the fluid can fluctuate between low-stress and high-stress  states.  Histograms and mean values were sampled from 12 long runs for each value of $\gd\Gamma_2$.
\label{fig04}
}
\end{figure}

\textbf{Inviscid transition.}
The $(\dot\gamma, L_y)$-parameter scans confirm the existence of low-viscosity phases when confinement geometry and shear-rate 
resonate with the natural bulk vortex size and circulation time scale of an active fluid (Fig.~\ref{fig03}b). The presence of an active driving mechanism is essential for the emergence of intrinsic length- and time-scales in  the statistically stationary non-equilibrium flow states~\cite{2013Dunkel_PRL}. It is therefore interesting to explore how a decrease in the activity, which can be realized experimentally through oxygen or nutrient depletion~\cite{2013Dunkel_PRL,2015LoGaDoAuCle}, affects the quasi-inviscid behavior. We study this process numerically through a systematic change of $\Gamma_2$, while keeping all other parameter fixed.  Starting from the low-viscosity state with $\Gc_2^*<0$ shown in Fig.~\ref{fig03}e, we increase  $\Gamma_2$ by adding an increment $\delta\Gamma_2>0$ to $\Gc_2^*$, corresponding to a decrease in activity.   As $\delta\Gamma_2$ increases,  the average viscosity undergoes a rapid increase before dropping to the value $\nu/\Gamma_0=1$  expected for a passive fluid with kinematic viscosity $\Gamma_0$ (Fig.~\ref{fig04}a). The viscosity peak separating the active from the passive phase can be explained by studying the stress distributions  (Fig.~\ref{fig04}b): Away from the transition region, the system remains locked in the quasi-inviscid or the laminar ground-state (blue and red curves in Fig.~\ref{fig04}b). In the critical transition regime, large fluctuations can cause the dynamics to oscillate between a low-stress ground-state and excited higher-stress states, resulting in a bimodal stress distribution and a higher average viscosity (green and orange curves in Fig.~\ref{fig04}b). 

%%%%%%%%%%FIG - superfluid transitions%%%%%%%%%%%%%
%%%%%%%%%%%%%%%%%%%%%%%

\begin{figure}[t!]
 \includegraphics[width=1\textwidth]{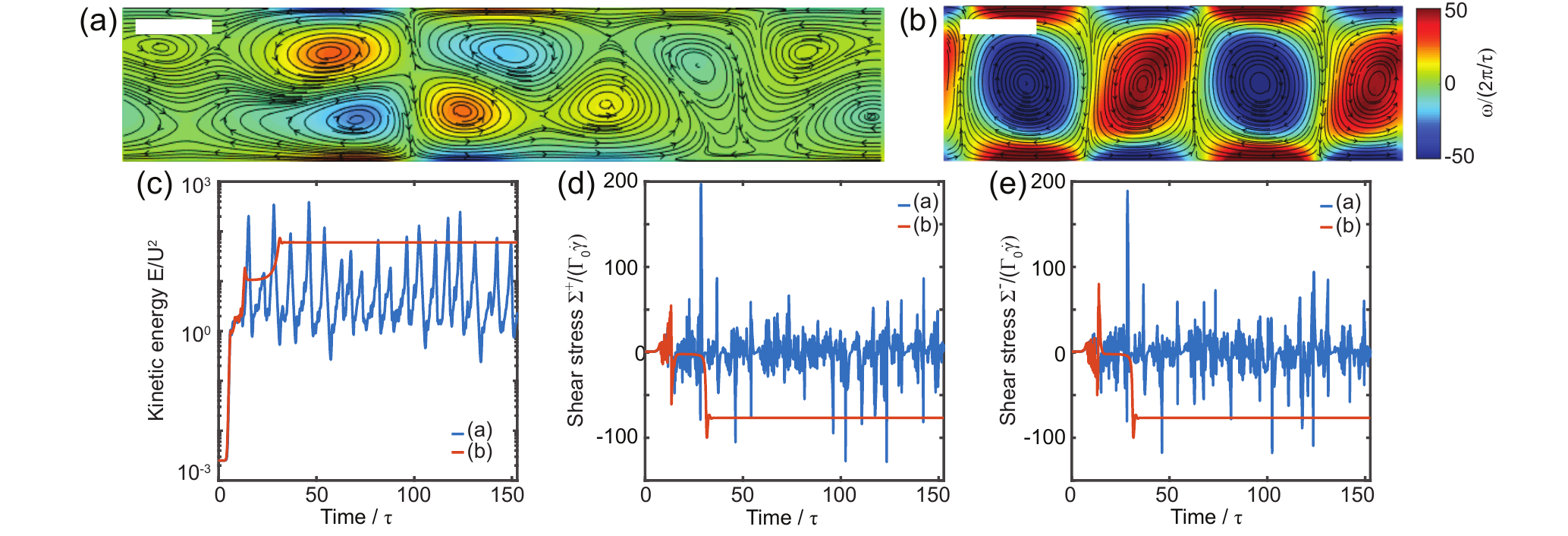}
 \caption{Decreasing the aspect ratio $\alpha=L_x/L_y$ stabilizes flow states capable of performing mechanical work. 
 (a-b), Steady-state flows for a narrow channel $L_y=2 \Gl$ and moderate shear $\dot{\gamma}=0.77\,\tau^{-1}$, shown for two different aspect ratios:   
 (a) $\alpha=5$ and  (b) $\alpha=3$.   
 (c), 
 The kinetic energy time series indicates that, for $\alpha=3$, the flow locks into a time-independent steady state, in which fluctuations are completely suppressed by the 
 no-slip shear boundary conditions.
 (d), 
Shear stresses $\Sigma^\pm(t)$ acting on the top (+) and bottom (--) boundary for the simulations in (a-b) yield a negative effective viscosity in both cases, implying that the fluid is pushing the boundaries.  This negative-viscosity effect is enhanced for the stationary state  observed at smaller aspect ratios (b).
\label{fig05}
 }
  \centering
   
\end{figure}

%\pagebreak
\textbf{Active fluids as motors.}
Work extraction from active suspensions has been investigated both theoretically~\cite{furthauer2012taylor,2015Kaiser} and experimentally~\cite{di2010bacterial,Sokolov19012010,PhysRevLett.112.158101} in recent years, resulting in a number of promising design proposals for bacteria-powered motors~\cite{2006Hiratsuka} and rectification devices~\cite{Galajda2007,2013Kantsler_PNAS}.  Moreover, recent experiments~\cite{2015LoGaDoAuCle} report long-lived  ($>25$s) negative viscosity flows in bacterial suspensions, supporting theoretical predictions that suggested the possibility of extracting work from polar active fluids~\cite{furthauer2012taylor}.  Equations~\eqref{e:eom} offer an alternative mechanism for constructing microbial \lq motors\rq{} by exploiting long-lived turbulent states that perform  work on the boundaries. Conditions for the existence of such states can be deduced analytically from energy balance considerations (App.~\ref{a:energy_balance}), which yields for the  power input 
\be
\label{e:energybalance}
P=\sum_{k}k^2(\Gamma_0+\Gamma_2 k^2+\Gamma_4 k^4)\varepsilon(k),
\ee
where $\varepsilon(k)$ is the energy spectrum at wavenumber $k$. For active fluids with $\Gamma_2<0$, the power input $P$ can become negative if the boundary conditions are tuned such that the energy spectrum  $\varepsilon(k)$ favors modes that produce a negative rhs. in Eq.~\eqref{e:energybalance}. Spectra of this type allow the extraction of mechanical work from the active fluid. We tested this idea by scanning different spectra $\varepsilon(k)$ through variation of the aspect ratio $\alpha=L_x/L_y$ of the simulation domain.  Our numerical results confirm the existence of long-lived work-performing states in the low-shear regime $V<U$ (Fig.~\ref{fig05}). In particular, when the aspect ratio is not too large,  $\alpha \sim 3$, and the boundary separation matches  twice the  bulk vortex scale, $L_y\sim2 \Gl$~(Fig.~\ref{fig05}b), then the active flow is found to lock into a stationary state, in which the shear forces exerted on the boundaries remain constant and have negative sign. In this case, a simple  active fluid motor is obtained by connecting the ends of the domain in Fig.~\ref{fig05}b, to form a cylindrical film. Such a setup could, in principle, be realizable with bacterial soap films~\cite{2009SkolovAranson}.

%%%%%%%%%%%%%%%%%%%%%%%
%%%%%%%%%%%%%%%%%%%%%%%

\textbf{Speculative quantum analogies.}
The periodic bulk solutions of Eqs.~\eqref{e:eomvsf} include inviscid vortex lattices (Fig.~\ref{fig01}b-d) reminiscent of those in quantum fluids~\cite{1957Abrikosov,1964Kleiner,1999lounasmaavortices,2012CheukZwierlein,2006bewleysuperfluid}. In particular, the lattice shown in Fig.~\ref{fig01}c is of Abrikosov-type (cf. Figures in~\cite{1957Abrikosov,1964Kleiner}). Of course, such marginally stable vortex lattices are exact solutions of our model only in a quasi-infinite fluid, and they become replaced by \lq cavity-modes\rq{} in the presence of confinement. However, lattice remnants remain visible in simulations with shear boundaries (Movie~1). Furthermore, the half-loops in the stress-field lines which form along the boundaries in the low-viscosity state (Fig.~\ref{fig03}e, bottom; Movie~4) bear a striking resemblance to the presumed edge-current structure in solid-state quantum Hall devices (cf. Fig.~1c in Ref.~\cite{2015Ketterle}). The role played by the stress tensor  for force transmission in an active fluid is comparable to that of the conductivity tensor for charge current transport in a quantum superfluid~\cite{Leggett_Book,1938Kapitza,1938AllenMisener}. The \lq superfluid\rq{} defects in the stress field of an active fluid reflect an interruption of force transmission lines between the boundaries giving rise to low-viscosity states (Fig.~\ref{fig03}e, bottom; Movie~4).  Apparent phenomenological similarities of active and quantum fluids can be traced back to the fact that  these two distinct classes of systems share two key features: (i) the governing equations describe collective low-energy excitations in the form of coherent vortex structures,  and (ii) unlike classical turbulence the emergent flow structures have a dominant length scale~\cite{2014Quantum_Turbulence}. In the quantum case, vortices are supported by an external magnetic field, whereas in active fluids vortices arise spontaneously from the microscopic and hydrodynamic interactions of bacteria~\cite{2012Sokolov,2013Dunkel_PRL}, ATP-driven microtubule bundles~\cite{2012Sanchez_Nature}  or other active components. Effective mathematical descriptions of quantum superfluids~\cite{RevModPhys.71.S318} build on Gross-Pitaevskii-type mean-field equations or semiclassical two-fluid models~\cite{1941Landau}.  In the future, it would be interesting to investigate whether, in some appropriate limit, such coarse-grained phenomenological  descriptions can be approximated by Eqs.~\eqref{e:eomvsf} through a suitably truncated Madelung transformation~\cite{Madelung1927} or by systematically eliminating one of the two velocity fields in two-fluid models.  Moreover, it will be interesting to explore whether biologically or chemically driven non-equilibrium flows described by Eqs.~\eqref{e:eom} and~\eqref{e:stress} can mimic other defining characteristics of conventional superfluids, such as wall-climbing Rollin films~\cite{1939RollinsSimon,1948Atkins} or the Hess-Fairbank effect~\cite{HessFaribank1967,Leggett_Book}.

\section*{CONCLUSIONS}
Phenomenological stress tensors of the type~\eqref{e:stress} provide a simplified description of
non-local stresses in non-Newtonian fluids~\cite{2014BipolarBook,1993BeNi_PhysD,1996Tribelsky_PRL,PhysRevE.77.035202}.  In pattern-forming liquids, such  higher-order stresses arise naturally from diagrammatic expansions~\cite{2011Ma}.  Although quantitatively more accurate stress tensors for complex active fluids likely include non-linear correction terms, it is expected that the generic long-wavelength expansion~\eqref{e:stress}  captures essential stability properties, similar to the success of Landau-type polynomial approximations for order-parameter potentials  in equilibrium phase-transition theories.  In particular, many pattern-forming liquids can be expected to have damped and growing modes that are separated by a zero-stress manifold in Fourier-space. Nonlinear advection and confinement can bias the flow dynamics towards spending substantial time periods in the vicinity of effectively frictionless states, suggesting that quasi-inviscid phases may be a quite generic feature of active fluids. If the predicted non-monotonic viscosity behavior in Fig.~\ref{fig03}b can be confirmed in future experiments, then the practical challenge reduces to designing fluids and confinement geometries that realize stress fields similar to that in Fig.~\ref{fig03}e.

\textbf{Acknowledgements.}
The authors thank Alex Townsend for advice on numerical questions, and Aden Forrow, Ruben Rosales and Francis Woodhouse for helpful discussions. This work was supported by an MIT Solomon Buchsbaum Fund Award (J.D.), an Alfred P. Sloan Research Fellowship~(J.D.) and an Edmund F. Kelly Research Award (J.D.). 

%%%%%%%%%%%%%%%%%
%%%%%%%%%%%%%%%%%
\appendix
\renewcommand\thefigure{\thesection.\arabic{figure}}
\section{Numerical Methods}
\label{numerical_methods}
%%%%%%%%%%%%%%%%%
%%%%%%%%%%%%%%%%%

We simulate typical shear experiments~\cite{2015LoGaDoAuCle} in which two parallel boundaries move in opposite directions, both at a constant speed $V$ (Fig.~\ref{fig02}a). 
After rescaling by $L_x/(2\pi)$ and $L_y/2$, the simulation domain is a rectangle $(x,y)\in \mcal{D}=[-\pi,\pi]\times [-1,1]$ with periodic boundary conditions in the $x$-direction and non-periodic conditions in the $y$-direction. The usual no-slip boundary conditions for the velocity field $\bs v$ translate into  $\p_y \psi(x,\pm 1)=\pm V-V_{\textnormal{CM}}$ and $\psi(x,\pm 1)=0$.

\par
A well-known challenge when working in the vorticity-stream function formulation is that the Poisson equation~\eqref{e:eomvsfb} is overdetermined by the combined Dirichlet [$\psi(x,\pm 1)=0$] and Neumann [$\p_y \psi(x,\pm 1)=\pm V-V_{\textnormal{CM}}$] boundary conditions for $\psi$. For the standard incompressible NS equations with no-slip boundary conditions this issue was resolved by Quartpelle and Valz-Gris \cite{Quartapelle1981}, who proposed to reinterpret the Neumann data for $\psi$ as a set of integral conditions for the vorticity $\omega$. In practice, the implementation of these integral conditions involves computing all the harmonic functions on a given domain. 
\par
To solve Eqs.~\eqref{e:eomvsf} numerically, we translate the integral conditions from the  corresponding classical Navier--Stokes problem, which specifies two boundary conditions.  Because Eq.~\eqref{e:eomvsfa} is a sixth-order PDE, we need four more constraints to determine the solution. We therefore additionally impose $\nabla^2\omega=0$ and $\nabla^4\omega=0$ at $y=\pm1$. This phenomenological choice corresponds to the assumption that the total force on the boundary coming from the higher-order terms (proportional to $\Gamma_2$ and $\Gamma_4$) vanishes in a rectangular geometry. Combined with the no-slip assumption, these  higher-order conditions suffice to close the system~\eqref{e:eomvsf}. 
\par
\begin{figure}[t]
  \caption{
 To estimate the effective viscosity at fixed separation $L_y$ and shear rate $\dot{\gamma}$ from an ensemble average, we generate $\geq 10$ simulations with initial data corresponding to a randomly perturbed linear shear profile (App.~\ref{numerical_methods}). 
  (a),~Time series of the kinetic energy $E(t)$ for multiple runs.
 (b),~Time series of the shear stress $\Sigma^+(t)$ on the upper boundary for $L_y=5 \Lambda$, $\dot{\gamma}=1.4\tau^{-1}$. The vertical dotted line indicates the relaxation time $T$. 
 (c),~Combined time series for $t>T$ from all runs of the shear stress $\Sigma(t)=\Sigma^{+}(t)+\Sigma^{-}(t)$ rescaled by the kinematic viscosity $\Gamma_0$ and shear rate $\dot{\gamma}$.
(d),~Histogram corresponding to the combined time series in (c) yields the estimates for the mean viscosity $\nu= \langle\Sigma \rangle/\dot{\gamma}$ and its variance.
(e),~Convergence of the mean viscosity estimates as a function of the averaging interval $\Delta$. 
(f), Relative magnitude of the Fourier-Chebyshev coefficients of the vorticity field at a random representative time of the simulation demonstrates geometric convergence to zero, confirming that the number of modes used in the simulations suffices to completely resolve  the dynamics at double precision accuracy $(\epsilon \sim 10^{-16})$.
\label{figS1}
}
\centering
\includegraphics[width=1\textwidth]{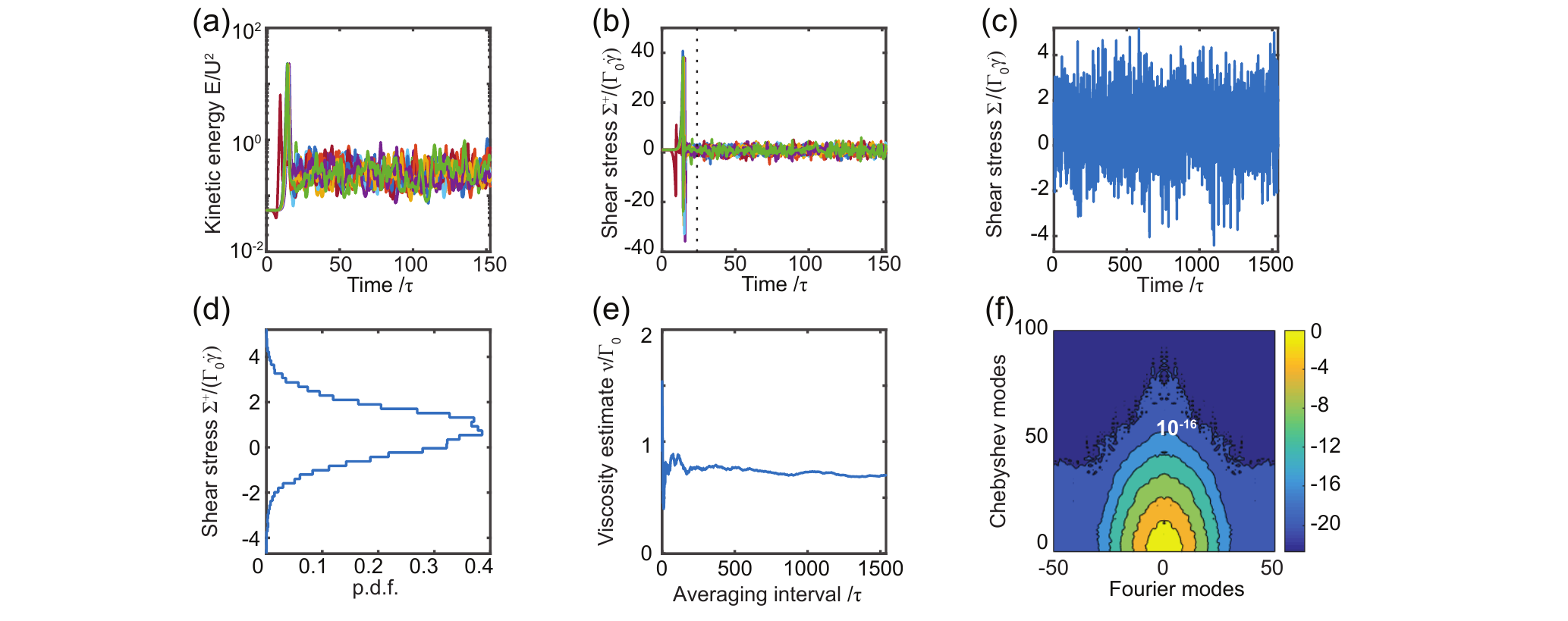}
\end{figure}

To evolve Eqs.~\eqref{e:eomvsf} in time, we use a third-order semi-implicit backward differentiation formula time-stepping scheme introduced by Ascher et al.~\cite{ascher1995implicit}, calculating the nonlinear advection term explicitly, while inverting the linear part implicitly. The instantaneous center-of-mass velocity is computed by integrating Eq.~\eqref{e:Newton}  with the forward Euler method. For the spatial discretization, we adopt a spectral method, expanding functions in a basis composed of Fourier modes and Chebyshev polynomials of the first kind. The implicit inversion is discretized using the well-conditioned scheme introduced by Olver and Townsend \cite{2013TownsendOlver,townsend2014automatic}. Since the system is periodic in the horizontal direction, the linear operator separates into one-dimensional operators, one for each Fourier mode. The resulting one-dimensional discretized linear operators augmented with integral and boundary conditions are sparse and almost banded, and therefore can be efficiently inverted. The explicit calculation of advection is done by collocation, that is, the relevant derivatives of $\omega$ and $\psi$ are evaluated on the Fourier--Chebyshev grid using the discrete Fourier transform (DFT) and the discrete cosine transform (DCT), then multiplied, and subsequently converted back to the expansion coefficients using the inverse DFT and the inverse DCT. Furthermore, the 3/2-zero-padding rule \cite{SpecMethodsFD} is applied during the explicit step, to ensure that no spurious terms arising from the finite discretization affect the collocation calculation. Advection is the most expensive part with a complexity of  $O(n \log n)$ when using the computationally optimal fast Fourier transform (FFT) for a discretization with $n=n_{\textnormal{C}} \times n_{\textnormal{F}}$, where $n_{\textnormal{C}}$ and $n_{\textnormal{F}}$ is the number of Chebyshev and Fourier modes, respectively. In our simulations, a discretization size of $n_{\textnormal{C}},n_{\textnormal{F}}\sim 10^2$ suffices to obtain geometric convergence to double-precision accuracy~(Fig.~\ref{figS1}f).
\par
Simulation runs are initiated as follows. For fixed shear rate $\dot{\gamma}$, a linear shear profile corresponds to a constant vorticity field $\omega_0=-2\dot{\gamma}$. We set $\omega=\omega_0+\textnormal{small noise}$, and then correct $\omega$ by projection so that it obeys the integral and boundary conditions. We then solve the Poisson Eq.~(\ref{e:eomvsfb}) for $\psi$. The such generated pair $(\omega,\psi)$ is then used to start the time-stepping scheme.
\par
Prior to scanning the parameter space relevant to the shear experiments, we validated our algorithm against  results obtained earlier~\cite{2015SlomkaDunkel} for the periodic case. When the separation between the boundaries is large compared to the vortex size, the effect of the boundaries becomes negligible, and we recover energy spectra consistent with those obtained for periodic boundary conditions as well as with corresponding analytical results. After this cross-validation, we applied the Chebyshev--Fourier spectral method to simulate shear experiments in active fluids. 

%%%%%%%%%%%%%%%%%%%%%
\section{Hodge decomposition}
\label{a:hodge}
%%%%%%%%%%%%%%%%%%%%%

In a two-dimensional planar region $\mcal{D}$ with boundary~$\p\mcal{D}$, the Hodge decomposition for a vector field $\bs v$ reduces to
\be
\label{eq:HodgeDecomposition2D}
\bs v=\nabla \phi+\nabla\wedge\psi+\nabla g+\bs H,
\ee
where $\phi$ and $\psi$ are scalar functions satisfying the boundary conditions $\phi|_{\p\mcal{D}}=\psi|_{\p\mcal{D}}=0$, $g$ is a harmonic function $\nabla^2g=0$ with arbitrary boundary data, and $\bs H$ is a harmonic vector field ($\nabla \cdot \bs H=0,\, \nabla\wedge \bs H=0$) that is tangential to the boundary $(\bs H^\perp=0)$.

\par
For divergence-free flow fields, Eq.~(\ref{eq:HodgeDecomposition2D}) simplifies to
\be
\bs v=\nabla\wedge\psi+\nabla g+\bs H,
\ee
because $\nabla \cdot\bs v=0$ makes $\phi$ harmonic with zero boundary data, implying $\phi=0$ throughout $\mcal{D}$. Moreover, imposing no penetration through the boundary ($\bs v^\perp=0$) fixes Neumann data for $g$ as $\bs n \cdot\nabla g=0$ on $\p \mcal{D}$, and therefore $g=$ constant throughout $\mcal{D}$. We are then left with
\be
\bs v=\nabla\wedge\psi+\bs H.
\ee
Given that $\psi$ vanishes on the boundary, the physical interpretation of the harmonic field $\bs H$ is that it accounts for the center-of-mass motion of the fluid. This follows from
\be
\label{eq:TotalMomentum}
\int_{\mcal{D}}\bs v=\int_{\mcal{D}}\bs H,
\ee
since $\int_{\mcal{D}}\nabla\wedge\psi$ vanishes because of $\psi|_{\p\mcal{D}}=0$.

%%%%%%%%%%%%%%%%%%%%%%%
%%%%%%%%%%%%%%%%%%%%%%%%%%%%%%%%%%%%%%%%

%%%%%%%%%%%%%%%%%%%%%%%
\begin{figure*}[t]
  \caption{
Additional flow examples for various channel widths $L_y$ and shear rates $\dot{\gamma}$. The shear stress histograms represent averages over $\geq$10 runs. 
\label{figS2}
}
\centering
\includegraphics[width=1\textwidth]{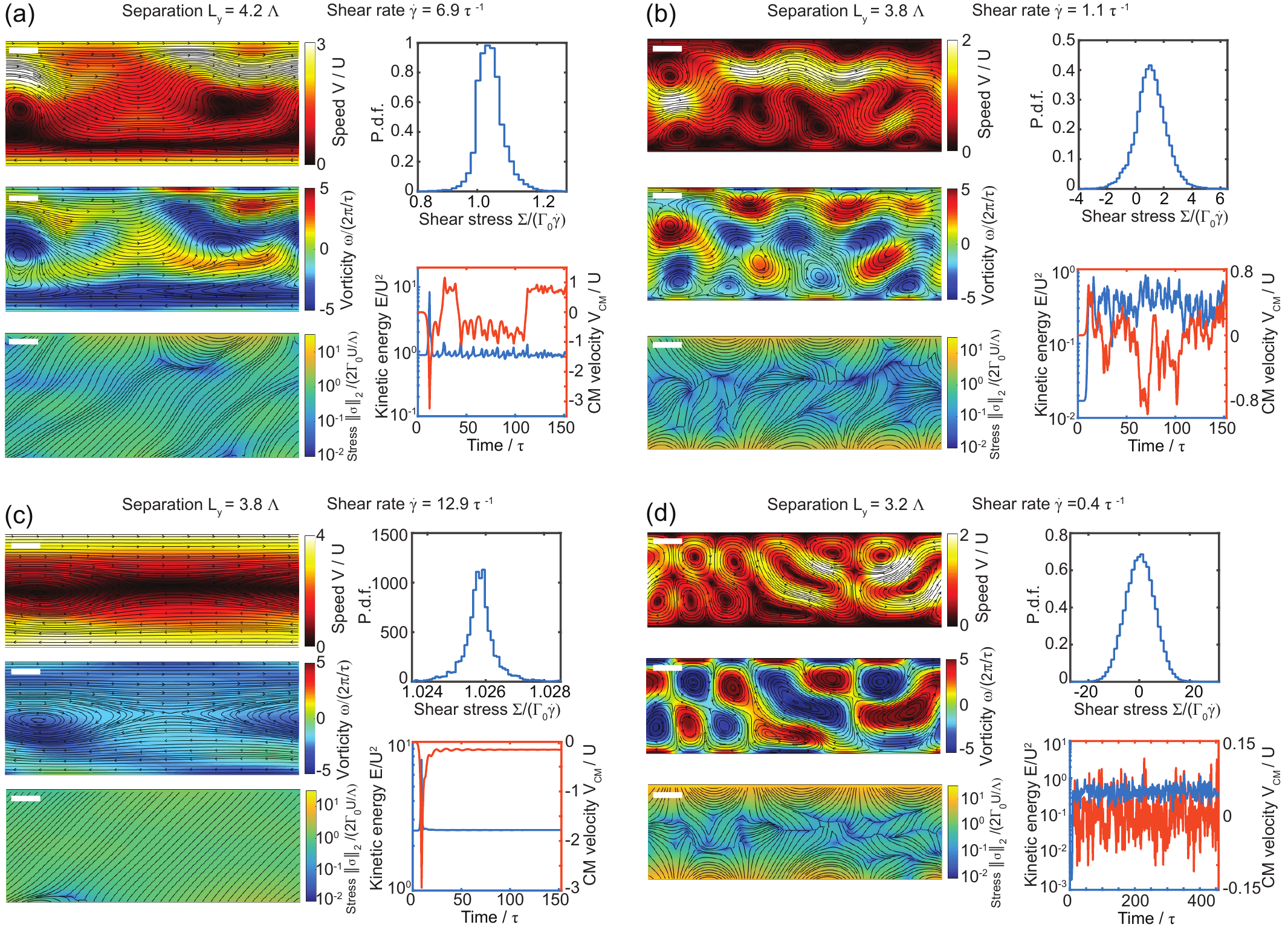}
\end{figure*}
%%%%%%%%%%%%%%%%%%%%%%%

Now consider a rectangle with periodic boundary conditions in the $x$-direction. We write the harmonic field as $\bs H=H_x \hat{\bs x}+H_y \hat{\bs y}$. Since $\bs H$ is harmonic, both $H_x$ and $H_y$ satisfy Laplace's equation. Additionally, $\bs H^\perp=0$ requires that $H_y=0$ on the boundary, and hence $H_y=0$ throughout the domain. The divergence-free condition, $\nabla\cdot \bs H=0$, requires that $H_x$ is a function of $t$ and $y$ only, $H_x(t,y)$. The curl-free condition, $\nabla\wedge \bs H=0$, further reduces $H_x$ to be solely a function of time. From Eq.~(\ref{eq:TotalMomentum}) we see that $H_x$ represents the center-of-mass speed,
\be
\bs H=V_{\textnormal{CM}}(t)\hat{\bs x},
\ee
where $\hat{\bs x}$ is the unit vector along $x$-axis. The dynamical equation for $V_{\textnormal{CM}}$ follows from Newton's Second Law
\be
M\f{dV_{\textnormal{CM}}}{dt}=F^{+}-F^{-},
\ee
where  $M$ is the total fluid mass, and $\bs F^+=F^+\hat{\bs x}$ and $\bs F^-=-F^-\hat{\bs x}$  are the forces on the upper and lower boundary (i.e., $F^{+}=F^{-}$ if the boundaries are pulled in opposite direction with equal force). Since $M=\rho L_x L_y$, where $\rho$ is the constant two-dimensional fluid density, we obtain
\be\label{e:Newton}
\f{dV_{\textnormal{CM}}}{dt}=\f{1}{L_y}(\Sigma^+-\Sigma^-),
\ee
where $\Sigma^{\pm}(t)=\f{1}{L_x}\int dx\,\sigma_{yx}(x,y=\pm 1)$ are the mean kinematic stresses as defined in the Main Text.

\par
The Hodge decomposition is also quite natural from an energetic perspective, for it provides an orthogonal splitting of the kinetic energy. In the present case, we have for the total kinetic energy,
\be
\mathcal{E}(t)&=&\f{1}{2}\int_{\mcal{D}}dxdy\, \bs v^2 
\notag\\ 
&=&\f{1}{2}\int_{\mcal{D}}dxdy\, [(\p_y\psi+V_{\textnormal{CM}})^2+(\p_x\psi)^2] 
\notag\\ 
&=&\mathcal{E}_{\psi}+\mathcal{E}_{\textnormal{CM}},
\ee
where the cross-term $\int_\mcal{D}dxdy\, \p_y\psi V_{\textnormal{CM}}$ vanishes by virture of  the boundary conditions imposed on $\psi$. Thus, the total kinetic energy splits into the vortical kinetic energy 
\be\notag
\mathcal{E}_{\psi}=\f{1}{2}\int_{\mcal{D}}dxdy\, \bs [(\p_y\psi)^2+(\p_x\psi)^2]
\ee 
and the center-of-mass kinetic energy 
\be
\notag
\mathcal{E}_{\textnormal{CM}}=\f{1}{2}\int_{\mcal{D}}dxdy\,V_{\textnormal{CM}}^2.
\ee 
Figure~3a shows the proportions of how the total kinetic energy splits between the two components.

%%%%%%%%%%%%%%%%%%%%%%%
\begin{figure}[t]
  \caption{
(a)~Validation that the flow symmetry breaking occurs is observed with equal probability for  both directions. Same parameters as  Fig.~1b-c ($L_y=2.2\Lambda$, $V=1.65U$). For 300 runs, we obtained 46.3:53.7 for the relative proportions of left-right symmetry breaking.  
(b)~Standard deviation of the effective viscosity shown in Fig. 3b. We distinguish between two regimes whose boundary (black line) is defined by the shear speed $V$ being equal to the characteristic vortex speed $U$. At small shear, $V<U$, the standard deviation is inversely proportional to the shear rate. That is, in the weak-shear regime, the fluctuations of the shear stress $\Sigma$ depend only on the channel width $L_y$. At large shear, $V>U$, the flow becomes more stable and the standard deviation quickly becomes orders of magnitude smaller than $\Gamma_0$. Blue lines indicate horizontal and vertical cuts shown in Fig. 3c,d.
\label{figS3}
\label{figS4}
}
\centering
\includegraphics[width=0.9\textwidth]{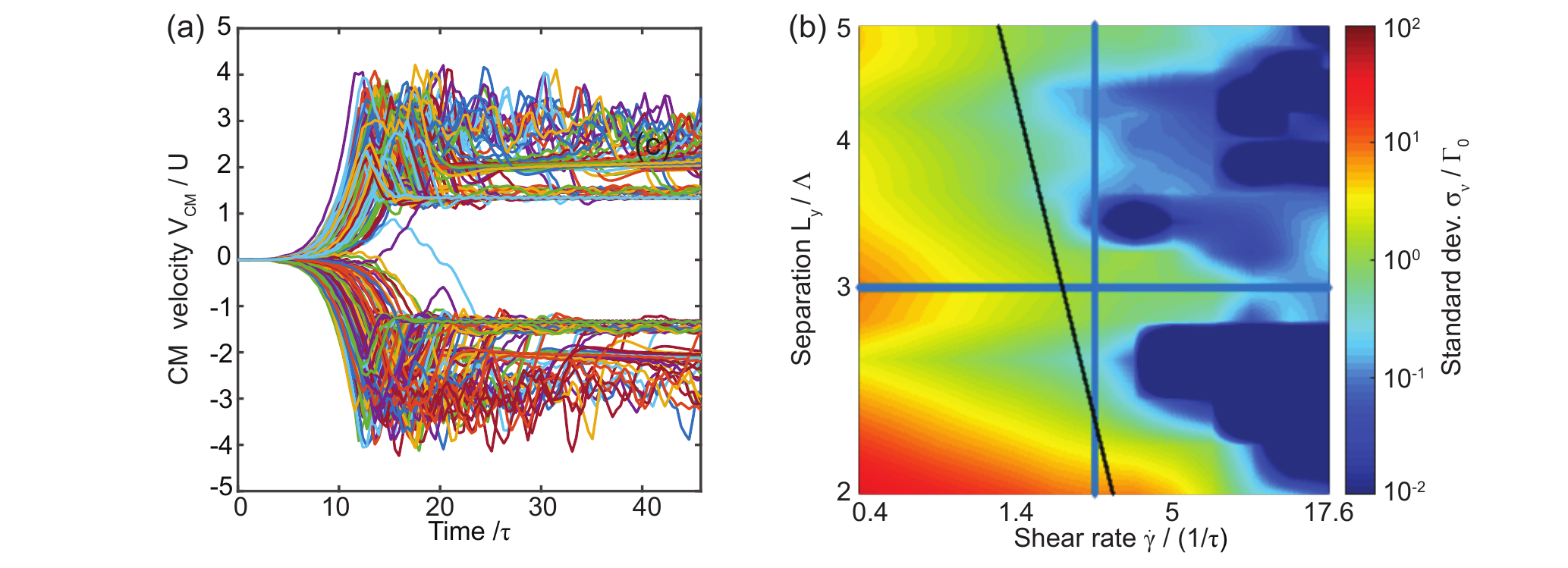}
\end{figure}
%%%%%%%%%%%%%%%%%%%%%%%

%%%%%%%%%%%%%%%%
\section{Characteristic scales}
\label{a:Characteristic scales}
%%%%%%%%%%%%%%%%

To derive characteristic length, time and velocity scales for the generalized Navier-Stokes model, consider the linearized vorticity equation
\be
\p_{t}\omega&=&\Gamma_0\nabla^2 \omega -{\Gamma}_2\nabla^4  \omega+{\Gamma}_4\nabla^6 \omega.
\ee
In Fourier space, this equation reads
\be
\p_{t}\hat{\omega}&=&-k^2(\Gamma_0+{\Gamma}_2 k^2+{\Gamma}_4 k^4 )\hat{\omega},
\ee
and has solutions of the form
\be
\hat{\omega}_k(t)=\hat{\omega}_k(0)e^{\sigma(k) t},
\ee
where $\sigma(k)=-k^2(\Gamma_0+{\Gamma}_2 k^2+{\Gamma}_4 k^4 )$. For $\Gamma_2<0$, the peak of the spectrum is well approximated by the maximum $k_p$ of the function $f(k)=\Gamma_0+{\Gamma}_2 k^2+{\Gamma}_4 k^4$, yielding 
\be
k_p^2=\f{-\Gamma_2}{2\Gamma_4}.
\ee
The associated wavelength is $\lambda_p=2\pi/k_p$. This wavelength represents two vortices, one with positive and one with negative vorticity, each of characteristic diameter
\be
\Lambda=\f{\lambda_p}{2}=\pi \sqrt{\f{2\Gamma_4}{-\Gamma_2}}.
\ee 
The corresponding growth rate is
\be
\sigma(k_p)&=&\notag
\f{\Gamma_2}{2\Gamma_4}\left(\Gamma_0-\f{\Gamma_2^2}{2\Gamma_4}+\f{\Gamma_2^2}{4\Gamma_4}\right)
\\
&=&
\f{\Gamma_2}{2\Gamma_4}\left(\Gamma_0-\f{\Gamma_2^2}{4\Gamma_4}\right),
\ee
which sets the time scale 
\be
\tau=\f{1}{\sigma(k_p)}.
\ee
If we roughly expect that, within time $\tau$, a fluid particle can travel around the vortex pair, then the characteristic speed is
\be
U
&=&
\pi \lambda_p \sigma(k_p) 
\notag\\
&=&
2\pi^2 \sqrt{\f{-\Gamma_2}{2\Gamma_4}}\left(\f{\Gamma_2^2}{4\Gamma_4}-\Gamma_0\right).
\ee

\section{Energy balance}
\label{a:energy_balance}
We derive the energy balance, Eq.~(8) of the Main Text, by considering  how the total kinetic energy, $\mathcal{E}(t)=\f{1}{2}\int_{\mcal{D}}dxdy\,\bs v^2$, changes with time (using an Einstein summation convention),
\be
\f{d{\mathcal{E}}}{dt}&=&\int_{\mcal{D}}dxdy\, v_i  \p_t  v_i\notag \\\notag
&=&\int_{\mcal{D}}dxdy\,  v_i (-v_j  \p_j v_i-\p_i p+\p_j\sigma_{ji}) \\\notag
&=&\int_{\mcal{D}}dxdy\,\Big\{-\p_i[v_i(\f{1}{2}v^2+p)]  +v_i\p_j\sigma_{ji}\Big\} \\\notag
&=&-\int_{\p\mcal{D}}dx\,[v_\perp(\f{1}{2}v^2+p)]+ \int_{\mcal{D}}dxdy\,v_i\p_j\sigma_{ji} \\
&=& \int_{\mcal{D}}dxdy\,v_i\p_j\sigma_{ji}.
\ee
In the second line, we used the equation of motion [Eq.~(1b)], in the third the incompressibility condition [Eq.~(1a)], in the fourth the divergence theorem ($v_\perp$ is the normal component to the boundary $\p \mcal{D}$), and in the last the fact that there is no penetration of the fluid through the walls ($v_\perp=v_y=0$ at $y=\pm 1$). Integration by parts further gives
\be
\f{d{\mathcal{E}}}{dt}&=& \int_{\mcal{D}}dxdy\,[\p_j(v_i\sigma_{ji})-(\p_j v_i) \sigma_{ji}]
\notag\\\notag
&=&\int_{\p\mcal{D}}dx\,\sigma_{yi}v_i-\int_{\mcal{D}}dxdy\,(\p_j v_i)\sigma_{ji} \\
&=&V(F^{+}+F^{-})-\int_{\mcal{D}}dxdy\,(\p_j v_i)\sigma_{ji}.
\ee
In the second line, we used the divergence theorem and in the last line the no slip boundary condition. $F^{+}$ and $F^{-}$ are the magnitudes of the force acting on the upper and lower boundaries, as defined above, and $V$ is the speed of the boundaries. We recognize $V(F^{+}+F^{-})$ as the power input, $P$, and therefore, for steady states with $d{\mathcal{E}}/dt=0$, we have
\be
P=\int_{\mcal{D}}dxdy\,(\p_j v_i)\sigma_{ji}.
\ee
Using the explicit form of the stress tensor [Eq.~(2) of the Main Text], we obtain
\be
P=\int_{\mcal{D}}dxdy\,(\p_j v_i)(\Gamma_0-\Gamma_2 \p_{nn}+\Gamma_4\p_{nn}^2)\p_{j}v_i.
\notag
\ee
In terms of Fourier modes, $\bs v=\sum_{\bs k}\hat{\bs v}(\bs k)e^{i \bs k \cdot \bs x}$, the balance reads
\be
P=\sum_{\bs k} k^2(\Gamma_0+\Gamma_2 k^2+\Gamma_4 k^4)|\hat{\bs v}(\bs k)|^2,
\ee
where $k=|\bs k|$. We now introduce the energy spectrum $\varepsilon(k)=\sum_{\bs k':|\bs k'|=k}|\hat{\bs v}(\bs k')|^2$ to recover Eq.~(7) of the Main Text
\be
P=\sum_{k} k^2(\Gamma_0+\Gamma_2 k^2+\Gamma_4 k^4)\varepsilon(k).
\ee

%%%%%%%%%%%%%%%%%%%%%%%%
%\newpage
%\bibliographystyle{unsrt}
%\bibliography{Jonasz_shear_combined}
%%%%%%%%%%%%%%%%%%%%%%%%

\end{document}